\documentclass[aps, prl, onecolumn, 10pt, noshowpacs, noshowkeys,  notitlepage, twoside, a4paper, superscriptaddress]{revtex4}
\usepackage{color}
\usepackage[utf8]{inputenc}
\usepackage{amsmath}
\usepackage{amsfonts}
\usepackage{amssymb}
\usepackage{graphicx}

\begin{document}
%
\title{Giant increase of diffusion by small rise of friction}
\author{Ivan G.  Marchenko}
\affiliation{NSC \lq\lq Kharkiv Institute of Physics and Technology\rq\rq, Kharkiv 61108, Ukraine}
\affiliation{Institute of Physics, University of Silesia, 41-500 Chorz{\'o}w, Poland}

\author{Igor I. Marchenko}
\affiliation{NTU \lq\lq Kharkiv Polytechnic Institute\rq\rq, Kharkiv 61002, Ukraine}

\author{Viktoriia Aksenova}
\affiliation{NSC \lq\lq Kharkiv Institute of Physics and Technology\rq\rq, Kharkiv 61108, Ukraine}
\affiliation{Karazin Kharkiv National University, Kharkiv 61022, Ukraine}

\author{Jerzy  {\L}uczka$^*$}
\email[Correspondence and requests for materials should be addressed to J.{\L}.]{ (e-mail: jerzy.luczka@us.edu.pl)}
\affiliation{Institute of Physics,  University of Silesia, 41-500 Chorz{\'o}w, Poland}
\author{Jakub Spiechowicz}
\affiliation{Institute of Physics, University of Silesia, 41-500 Chorz{\'o}w, Poland}
\begin{abstract}
\noindent  {\bf Abstract} \\
Diffusion coefficient usually decreases when friction increases. We analyze the opposite behavior in the paradigmatic system consisting of an inertial Brownian particle moving in a symmetric spatially periodic potential and driven by an unbiased time periodic force. For tailored parameter set in strong dissipation regime the particle spreading can be giantly amplified: if the friction is twice as large then the diffusion grows up to five orders of magnitude. The mechanism lying behind this effect is related to bifurcation of periodic orbits oscillating  around the potential maximum and their symmetric displacement towards the adjacent potential minima when the friction coefficient increases.
On the other hand, in the weak dissipation regime, where the increase of diffusion vs friction  is also observed, the effect is induced by a non-monotonic change of population of the running orbits. However, in this regime the enhancement of diffusion is much smaller. 
\end{abstract}
\maketitle

\noindent {\Large {\bf Introduction}} \\
The celebrated Einstein relation for the diffusion coefficient $D$ of the force-free Brownian particle is given by the following relationship \cite{einstein,entropy23} 
\begin{equation}
D = D_E = \frac{k_B T}{\gamma}, 
\end{equation}
where $k_B$ is the Boltzmann constant, $T$ is temperature of the medium (thermostat) in which the particle moves and $\gamma$ is the friction (damping) coefficient which characterizes the interaction strength of the Brownian particle with the environment. The same relation is also fulfilled for a Brownian particle subjected to a constant force $F={const}$. From this equation it follows that $D_E$ depends inversely on the friction coefficient $\gamma$. In other words, $D_E$ decreases when $\gamma$ increases. 
Studies on e.g. surface diffusion reveals that in some regimes the more general relation is satisfied, namely, 
\begin{equation}
\label{invers}
D \sim \frac{1}{\gamma^{\sigma}}.  
\end{equation}
The exponent $\sigma$ is not universal but depends on a system, its model and the corresponding assumptions. Different values have been reported like $\sigma =1/2$ \cite{sigma05}, $\sigma = 1/3$ \cite{sigma03}, $\sigma \in (0.64, 0.76)$ \cite{sigma064}, $\sigma \in (0, 1/3)$  \cite{sigma033}.  Non-standard behavior of $D$ on the friction coefficient $\gamma$ is also possible. A good example is Brownian diffusion of skyrmions within magnetic materials \cite{skyrmion1}. For vanishing topological charge, $D$ monotonically decreases with growing $\gamma$. However, if the topological charge is non-zero, the diffusion coefficient increases until it reaches its maximum and next it decreases when the friction constant grows \cite{skyrmion2}. In this paper, we study the issue of non-monotonic dependence of diffusion on the friction coefficient for the Brownian particle moving in a spatially periodic potential and driven by a time-periodic force, i.e., the model which is relevant to a variety of different context \cite{risken}, including pendulums \cite{gitterman}, Josephson junctions \cite{kautz1996}, superionic conductors \cite{fulde}, dipoles rotating in external fields \cite{coffey}, Frenkel-Kontorova models \cite{frenkel}, charge density waves \cite{gruner} and cold atoms dwelling in optical lattices \cite{renzoni, denisov}, to name only a few.  
We mention that a similar system, however with an additional constant force, has been studied in Ref. \cite{march-prl}. 

Below we describe in detail the model of the Brownian particle moving in a spatially periodic potential driven by an unbiased time-periodic force and analyze the diffusion coefficient as a function of dissipation, i.e. how $D$ depends on the friction coefficient $\gamma$. We show that there are several distinct regimes in which (i) $D$ monotonically decreases with respect to $\gamma$ or (ii) $D$ is non-monotonic and exhibits a maximum as a function of $\gamma$. 
Within a tailored parameter set we identify a giant enhancement of the diffusion coefficient when the friction coefficient is slightly increased. We explain the mechanism responsible for this phenomenon. We also apply the analytical but approximate theory of vibrational mechanics and verify its relevance to the studied problem. \\

\noindent {\Large {\bf Model of Brownian dynamics}} \\
\noindent   We study diffusion properties of the Brownian particle moving in a spatially periodic structure and subjected to a time-periodic force. The corresponding Langevin equation assumes the dimensionless form \cite{marchPRE23}
\begin{equation}
	\label{La}
	\ddot{x} + \gamma\dot{x} = -\sin{x} + a \sin (\omega t + \phi_0) +  \sqrt{2\gamma Q} \, \xi(t).
\end{equation}
Note that in this scaling the dimensionless mass $m = 1$. The parameter $\gamma$ is the friction coefficient and $Q \propto k_B T$ is the dimensionless temperature of the system. The coupling of the particle with thermostat is modeled by the $\delta$-correlated Gaussian white noise $\xi(t)$ of zero mean and unit intensity, i.e. $\langle \xi(t) \rangle = 0$ and \mbox{$\langle \xi(t)\xi(s) \rangle = \delta(t-s)$}. The starting dimensional equation is presented in the section Methods, see Eq. (\ref{model1}),  where the corresponding scaling and dimensionless parameters are defined. 

The complexity of the underlying dynamics with the nonlinear force  $f(x) = -\sin{x}$ and the corresponding spatially periodic potential $W(x)= -\cos(x)$, the time-periodic force $g(t)= a\sin(\omega t+\phi_0)$ of the amplitude $a$, frequency $\omega$, initial phase $\phi_0$ and  thermal fluctuations of the intensity $2\gamma Q$  do not allow for an analytical approach which presently is clearly beyond the known mathematical methods. Therefore numerical simulations have been employed \cite{spiechowicz2015cpc}. 

In the deterministic case, when $Q = 0$, the system may be non-ergodic \cite{spiechowicz2016scirep} and therefore sensitive to the specific choice of the  initial conditions: position $x(0)$, velocity $v(0)$ of the particle and the initial phase $\phi_0$ of the ac-driving. Therefore all quantities characterizing the system  should be averaged over $\{x(0), v(0), \phi_0\}$ with uniform distributions to get rid of this dependence. However, for any non-zero temperature $Q>0$ the system is ergodic and the initial conditions do not affect its properties in the long time stationary regime. Therefore the mean value of observables can be accurately estimated by averaging any realization over a sufficiently long time interval. \\

\noindent {\Large {\bf Results: Impact of friction on diffusion } }\\
Despite its apparent simplicity, the system (\ref{La}) is extremely versatile, exhibits an impressive diversity of behavior and can successfully describe a number of important and various  physical situations. We will be interested only in one aspect of these properties, namely,  diffusion and its dependence on dissipation characterized by the friction coefficient $\gamma$.  Because the parameters space  $\{\gamma, a, \omega, Q\}$ is four dimensional we consider only the selected parameter regime in which the diffusion process in the long time limit is normal and described by a time-independent diffusion coefficient $D$. Hence it can be defined as \cite{jpc}
\begin{equation}
	\label{D}
	D =  \lim_{t \to \infty} \frac{1}{2t}\, \langle \left[x(t) - \langle x(t) \rangle \right]^2 \rangle.
\end{equation}
%
%

\begin{figure}[t]
\centering
\includegraphics[width=0.45\linewidth]{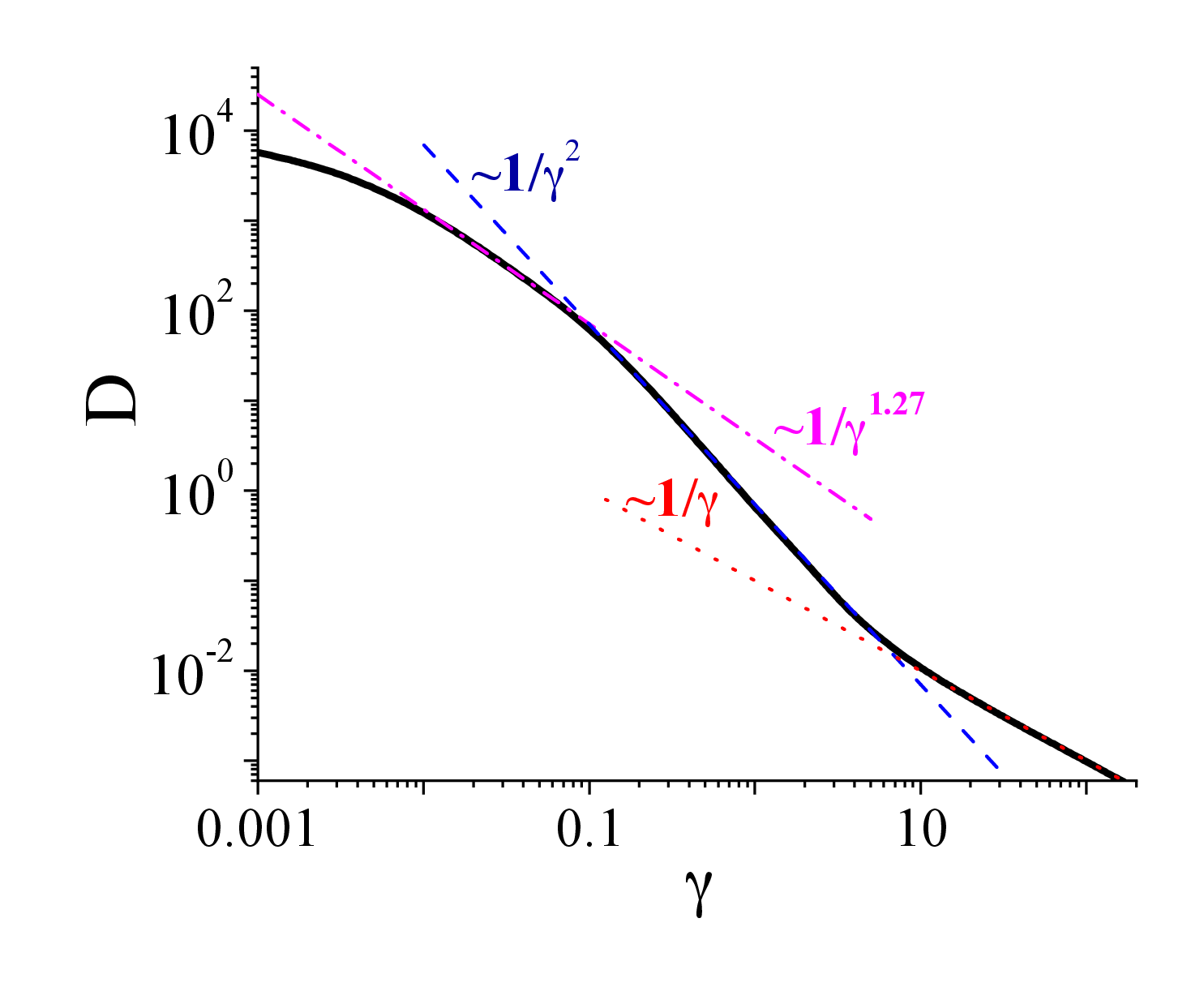}
\caption{From the weak to strong dissipation regime: The diffusion coefficient $D$ versus the  friction coefficient $\gamma$. Parameters are: the ac-driving amplitude $a=6$,  its  frequency 
$\omega = 1.59$ and  temperature  $Q=0.5$. }
\label{fig1}
\end{figure} 
The exact expression for the diffusion coefficient is not known for the model (\ref{La}) and only numerical results are available. In Ref. \cite{march-chaos} it was shown that for a tailored parameter regime the diffusion coefficient $D$ displays the damped quasi-periodic dependence on the amplitude $a$ of the time-periodic force $g(t) = a\sin{(\omega t +\phi_0)}$. 
The origin of this phenomenon was explained in Ref. \cite{marchPRE23} and is rooted in the complex deterministic structure of the nonlinear dynamics governed by a variety of stable and unstable orbits. 
The dependence of the diffusion coefficient $D$ on the friction coefficient $\gamma$ has not been studied in Ref. \cite{marchPRE23}. 
In Fig. 1 we show expected decrease of  $D$ versus $\gamma$ for a set of some fixed parameters. For the driving amplitude $a=6$ and temperature $Q=0.5$  one can distinguish three distinct intervals of functional dependence: (i) for friction $\gamma \in (0.01, 0.1)$ the diffusion coefficient diminishes like $1/\gamma^{1.27}$; (ii) for $\gamma \in (0.1, 8)$, it decays like $1/\gamma^2$; (iii) for larger $\gamma$ it again decreases like $1/\gamma$. The above exemplified monotonic dependence of $D$ on $\gamma$ occurs in extended domain of the parameter space.  However, this picture is changed radically when temperature $Q$ is significantly lowered.  
We have searched a part of the three dimensional parameter space $\{\gamma, a, Q\}$  in order to locate regions of non-monotonic dependence of $D$ on $\gamma$. 


We remind that in the noiseless system $Q=0$ corresponding to Eq. (\ref{La}), the dynamics may become complex, exhibiting both regular and chaotic behavior \cite{kissner} in the four-dimensional parameter space $\{\gamma, a, \omega, Q\}$ in which running and localized states (periodic orbits) can coexist \cite{marchPRE23}. The particle can oscillate in the periodic potential wells and next proceed either forward or backward within one or many temporal periods, or it can permanently move in one direction. Depending upon the system parameters and initial conditions various stationary regimes can be described by a rough approximation
\begin{equation}
\label{solut}
	x(t)  =  x_c +  G(\omega t) \pm k\omega t,  
\end{equation}
where $G(\omega t)$ is a periodic function of time and characterizes oscillations,  the parameter $x_c$ is a center of  oscillations and  $\omega$ is the dimensionless frequency of the time periodic force. In the case of locked trajectories $k=0$ and for running solutions $k \neq 0$. For the mechanical pendulum, $k=0$ corresponds to oscillations around $x_c$. The case $k \ne 0$ corresponds to rotation of the pendulum. In our recent work \cite{marchPRE23} we exemplified the system parameter sets for which locked and running solutions can emerge. Generally, it is difficult to establish a relationship between values of the parameters and initial conditions to answer the question for which of their subsets running or locked trajectories can be observed. However, our intuition suggests that for a given parameter set and for sufficiently large $\gamma$ there are only locked states.

Due to symmetry of the system (\ref{La}), in the long time limit the directed velocity $\langle v \rangle$ must vanish for both zero and non-zero temperature regimes, 
\begin{equation}
	 \langle v \rangle = 0. 
\end{equation}
Here and below $\langle \cdot \rangle$ stands for the average over the initial conditions and the ensemble of thermal noise realizations. In the deterministic case $Q=0$ for each trajectory  (\ref{solut}) with $k>0$ there is a set of initial conditions for which the partner trajectory with the value $-k$ travels in the opposite direction.  
 
For the noisy system with $Q>0$, there are several types of contribution to the spread of trajectories and in consequence to the diffusion coefficient $D$. The first, which is the largest one, comes from the distance between the running trajectories moving into the opposite directions. The second, also relatively large, is the spread between the running and periodic orbits. The next follows from contributions driven by thermal fluctuation in each of the running and localized states. Clearly, even at high temperature the latter are remarkably smaller than the two previous ones. Therefore it is intuitively expected that the diffusion coefficient $D$ will change considerably especially when the population of running and locked states is significantly modified. An example of various types of trajectories is  visualized in Fig. 4 in Ref. 
\cite{marchPRE23}.  \\

\begin{figure}[t]
\centering
\includegraphics[width=0.5\linewidth]{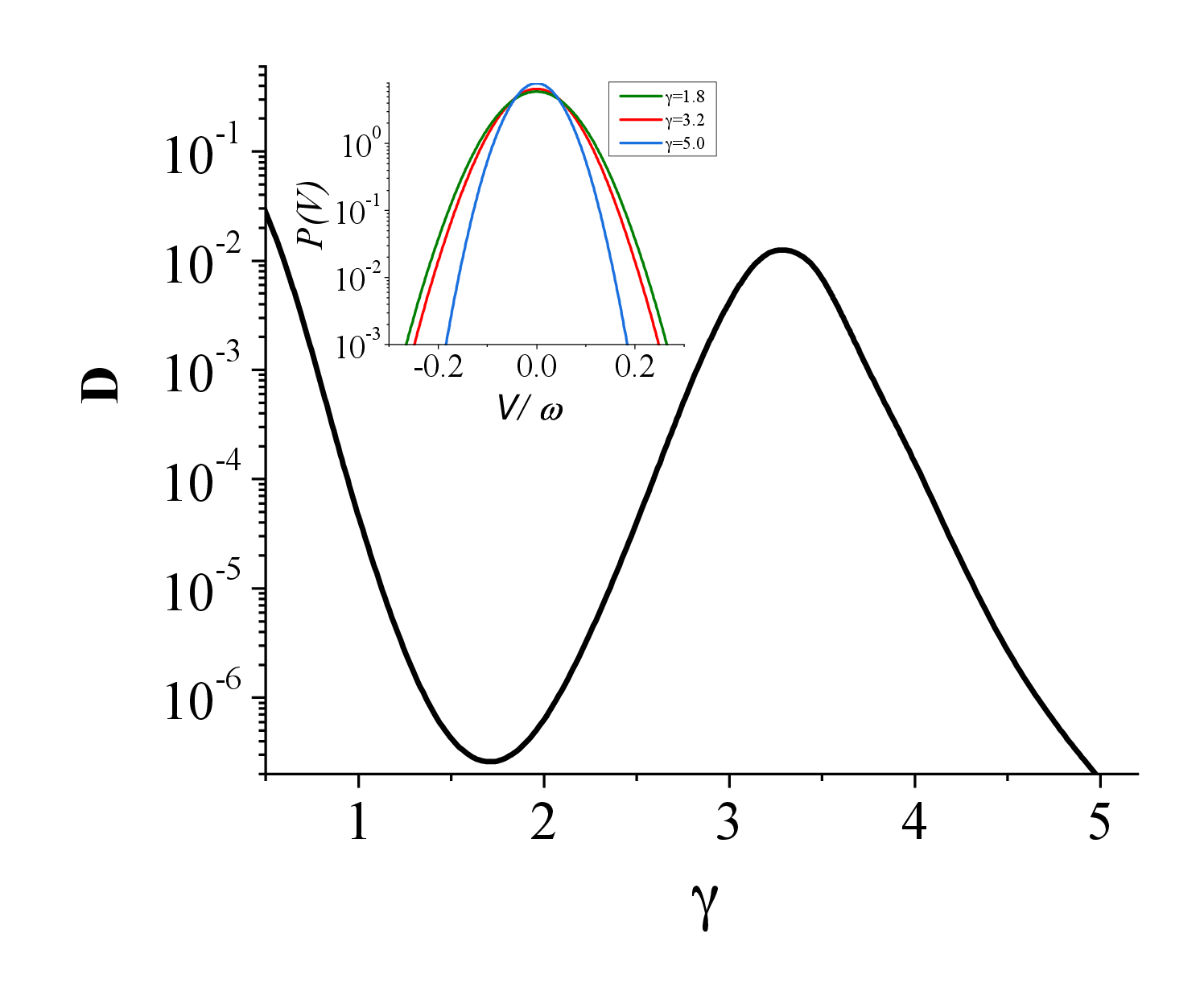}
\caption{The strong dissipation regime: Non-monotonic dependence of the diffusion coefficient $D$ on the friction coefficient $\gamma$. Inset: The probability distribution $P(V)$ for the period-averaged velocity $V$ of the Brownian particle.  
Parameters are: the ac-driving amplitude $a=14$, its frequency $\omega = 1.59$ and temperature  $Q=0.05$.
 }
\label{fig2}
\end{figure}

\noindent {\bf A. Giant enhancement of diffusion}\\
The case illustrated in Fig. \ref{fig1} is in fact only a quantitative modification of the monotonic decay $D \sim 1/\gamma$. Much more interesting is the non-monotonic dependence $D(\gamma)$. Such a situation is a {\it qualitatively new phenomenon}: the non-monotonicity implies the occurrence of local extrema, i.e. maxima and/or minima. An example of this behavior is visualized in Fig. \ref{fig2}. For the driving amplitude $a=14$, its frequency $\omega = 1.59$ and lower temperature $Q=0.05$, the diffusion coefficient decreases for small values of $\gamma$ and attains the local minimum $D=2.5 \times 10^{-7}$ at $\gamma = 1.66$. Next, it starts to increase to the local maximum $D = 1.4\times 10^{-2}$ at $\gamma =\gamma_M \approx 3.3$. Further increase of $\gamma$ results in a monotonic decline of $D$ to zero. We note that the enhancement of $D$ is $5.6 \times 10^{4}$, i.e. almost by five orders of magnitude when friction is increased less than twice. The main question is why $D$ starts to increase  when the friction grows and whether this behavior is exception or can occur in a wider domain of the parameters space. We shall now explain the origin of this phenomenon. 
In the deterministic counterpart of the noisy system  there are localized states with  $V = 0$  and running states with $V = \pm k \omega, k=1, 2, 3, ...$. The period averaged velocity is defined as 
\begin{equation}
	V = \lim_{t \to \infty} \frac{1}{\mathsf{T}} \int_t^{t+\mathsf{T}} ds \, \dot{x}(s), \quad \mathsf{T}=\frac{2\pi}{\omega}. 
\end{equation}

\begin{figure*}[t]
	\centering
	\includegraphics[width=0.32\linewidth]{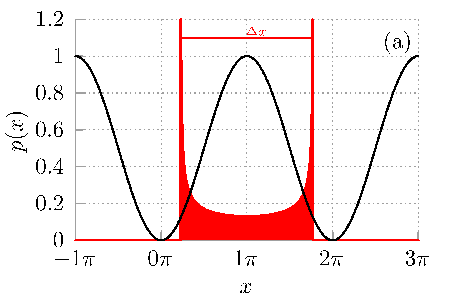}
	\includegraphics[width=0.32\linewidth]{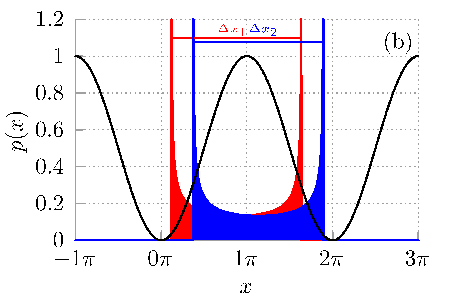}
	\includegraphics[width=0.32\linewidth]{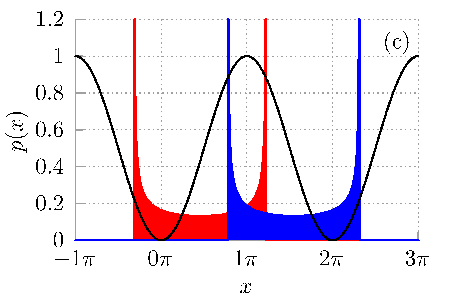}\\
	\includegraphics[width=0.32\linewidth]{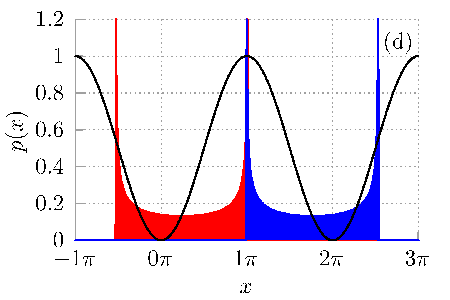}
	\includegraphics[width=0.32\linewidth]{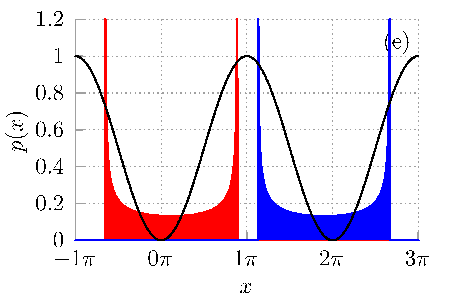}
	\includegraphics[width=0.32\linewidth]{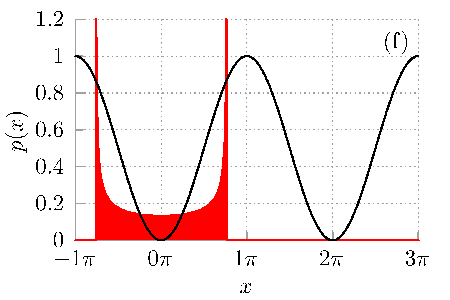}\\
	\includegraphics[width=0.32\linewidth]{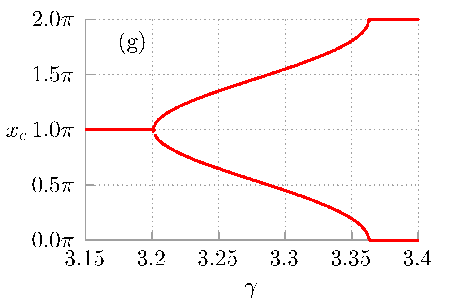}
	\caption{The probability density $p(x)$ for  position of the particle in the long time regime averaged over the initial conditions $x_0$, $v_0$ and $\phi_0$. The system is deterministic $Q = 0$.  Black curves: the potential $W(x) = -\cos{x}$.  Other parameters are: $a = 14$, $\omega = 1.59$ and the friction coefficient: $(a) \, \gamma = 3.15;$ $(b) \, \gamma = 3.21;$ $(c) \, \gamma = 3.3;$ $(d)\,  \gamma = 3.345;$ $(e) \, \gamma =3.36;$ $(f) \, \gamma =3.41$. 
For $\gamma > 3.2$, in dependence of initial conditions, two classes of periodic orbits emerge  which are symmetrically shifted 
in the positive (blue)  and negative (red) directions from the value $x_M=\pi$.	
	Bottom:  The pitchfork bifurcation: if $\gamma < 3.2$ the centers $x_c$ of periodic orbits are located at the maximum $x=\pi$ of the potential 
	$W(x)$  and for  $\gamma > 3.2$ they are shifted to its adjacent minima $x=0$ and $x=2 \pi$. }
	\label{fig3}
\end{figure*}
In the insets of Fig. \ref{fig2} the probability distribution $P(V)$ for the period-averaged velocity $V$ of the Brownian particle is depicted for the detected  maximum of the diffusion coefficient $D$. In the regime of strong damping  $\gamma \sim 3$ the running states are short lived and the period-averaged velocity distribution $P(V)$ exhibits only one maximum corresponding to the locked orbits ($k=0$ in Eq. (\ref{solut})). This fact in turn is important for explanation of the giant enhancement of $D$. We first look at the probability density $p(x)$ for the position of the particle in the long time regime. In Fig.~\ref{fig3} we present this quantity for the deterministic trajectories $Q = 0$ averaged with respect to the initial position $x_0$, velocity $v_0$ and phase $\phi_0$ distributed uniformly over the intervals $[0, 2\pi]$, $[-2,2]$ and $[0, 2\pi]$, respectively. We note that the averaging over the phase $\phi_0$ in the long time regime is equivalent to the averaging over the period of the external driving. For the convenience of the reader in the same plot we additionally illustrate the periodic potential $W(x) = -\cos{x}$. The reader can observe the characteristic {\it arc-sine} form of $p(x)$ \cite{levy,arcsine} which here is a consequence of averaging over the phase $\phi_0$.

We start our consideration assuming the value of $\gamma = 3.15 < \gamma_M$: if $\gamma$ increases then $D$ increases. It turns out that in the presented regime  (panel (a) in Fig. \ref{fig3}) the deterministic particle symmetrically oscillates around the \emph{maximum} $x = x_M = \pi$ of the potential and covers the indicated distance $\Delta x$ between its minimal and maximal position. The quantity $\Delta x/2$ can be interpreted as the amplitude of oscillations and the boundary values of this interval correspond to the most probable values of the particle position. In this case the center of oscillations is $x_c=\pi$ and oscillations around the maximum of the potential (the inverted pendulum) are not astonishing taking into account the large amplitude $a=14$ of the time periodic force. Note that the total "rocking" potential $\mathbb{W}(x, t) = -\cos(x) - a x \cos(\omega t+\phi_0)$ has local extrema (and the barriers) only when $a|\cos(\omega t+\phi_0)| < 1$. Therefore  for the amplitude $a=14$ in majority of time intervals ($95.5\%$ of full time) there are no barriers. Only in short time intervals ($4.5\%$ of full time) the barriers emerge and the potential is indeed the washboard one. Two extreme forms $\mathbb{W}_1 = -\cos{x}-14x$ and $\mathbb{W}_2 = -\cos{x} +14x$ looks like straight lines, the barriers disappear for some time intervals and the particle can "dance" around any position. 

If now $\gamma$ grows to the value $\gamma_1 = 3.2$, the amplitude of oscillations decreases and the spatial interval $\Delta x$ around the maximum $x_M$ of the potential $\mathbb{W}(x)$ decreases. In consequence,  in this window the effective potential barrier, the particle needs to overcome to diffuse due to thermal agitation, is reduced as well. It explains why the diffusion coefficient $D$  can increase despite the growth of friction $\gamma$.  We should  also remember that in the noisy system the intensity $\gamma Q$ of thermal noise $\xi(t)$ also increases as $\gamma$ increases and it is the second factor which can lead to the increase of $D$. 

Next, at $\gamma_1 = 3.2$ the "phase transition" arises: In dependence of initial conditions, two classes of periodic orbits emerge with two values of $x_c = x_{c1}$ and $x_c = x_{c2}$ which are symmetrically shifted 
in the positive and negative directions from the value $x_M=\pi$. When $\gamma$ increases  $x_{c1}$ and $x_{c2}$ move towards $x=0$ and $x=2\pi$, respectively,  see panels (b) and  (c) in Fig. \ref{fig3}.  These oscillations are not symmetric around $x_M$ but their centers are shifted towards the potential minima. It means that one of the turning points $x_e$ of the orbit (where the particle velocity changes sign) is closer to the potential maximum and the barrier height is further reduced, see the two inner vertical blue and red lines in panel (c) of Fig. \ref{fig3}.   

For $\gamma_M = 3.345$ one of the turning points of both blue and red orbits is exactly at the potential maximum $x_M$ and consequently the barrier height is zero, cf. panel (d) in Fig. \ref{fig3}. When the particle is approaching $x_M$ its velocity decreases to zero and for a relatively long time it resides in the neighborhood of $x_M$. In consequence, thermal fluctuations can freely spread the particle. 

For the friction coefficient lying in the interval $\gamma \in (3.345, 3.365)$ one observes two periodic orbits related to non-symmetrical oscillations around the potential minima for which the effective barrier height is greater than for $\gamma_M$ and therefore the diffusion coefficient starts to decrease as it is expected, see panel (e) in Fig. \ref{fig3}. At $\gamma_2 = 3.365$ the reversed bifurcation is detected: two previously observed locked trajectories now symmetrically oscillate around two  neighbouring  potential minima and  can be regarded as the same by a shift of the blue orbit to the left,  see panel (f) in Fig. \ref{fig3}. 

For $\gamma > \gamma_2$ we observe a decrease in the spatial interval $\Delta x$ of oscillations around the minimum. In other words, oscillations are focused closer and closer to the minimum of the potential as $\gamma$ grows. As a result, the effective potential barrier height increases and the diffusion coefficient decreases. To sum it up, the sequence of panels in Fig. \ref{fig3} reveals the mechanism which is responsible for the non-monotonic behavior of diffusion coefficient $D$ in the strong damping regime. We should be aware of limitations of this picture because in the noisy system  its
dynamics is much more complicated. 

In panel (g) of Fig. \ref{fig3} we collect the above information and extend it to continuous change of $\gamma$. It has the form of the bifurcation diagram for the centers of periodic orbits $x_c$  as a function of the friction coefficient $\gamma$.  The crucial point is emergence of two classes of periodic orbits for $\gamma > 3.2$.  For the first class the centers $x_c$ move to the minimum $x=0$ of the potential $W(x)$ whereas for the second $x_c$ goes to its another minimum $x=2 \pi$. It is visualized as a pitchfork bifurcation and reveals the mechanism which is responsible for the non-monotonic behavior of diffusion coefficient $D$ versus dissipation in the strong damping regime. \\

\begin{figure}[t]
\centering
\includegraphics[width=0.45\linewidth]{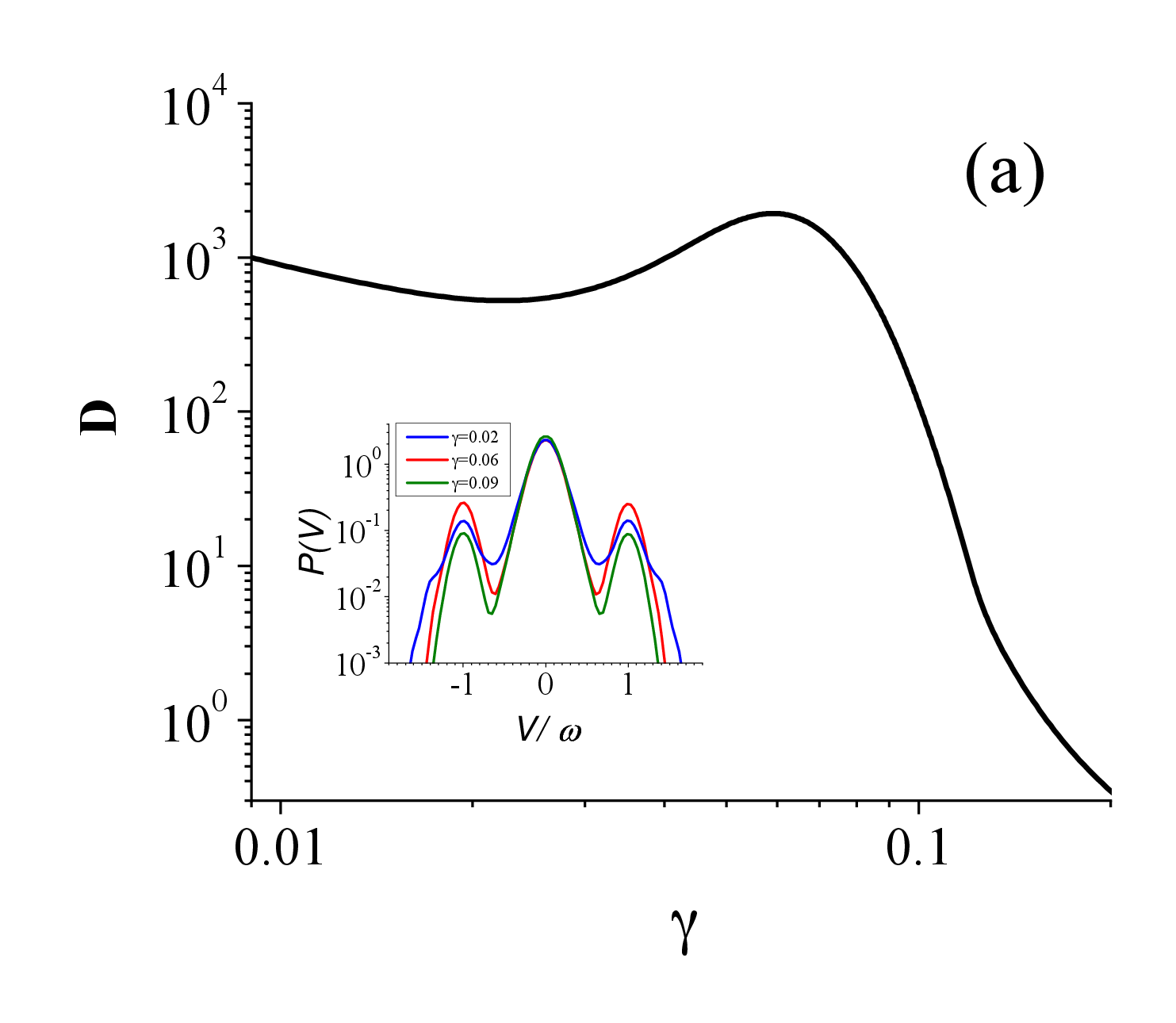}
\includegraphics[width=0.45\linewidth]{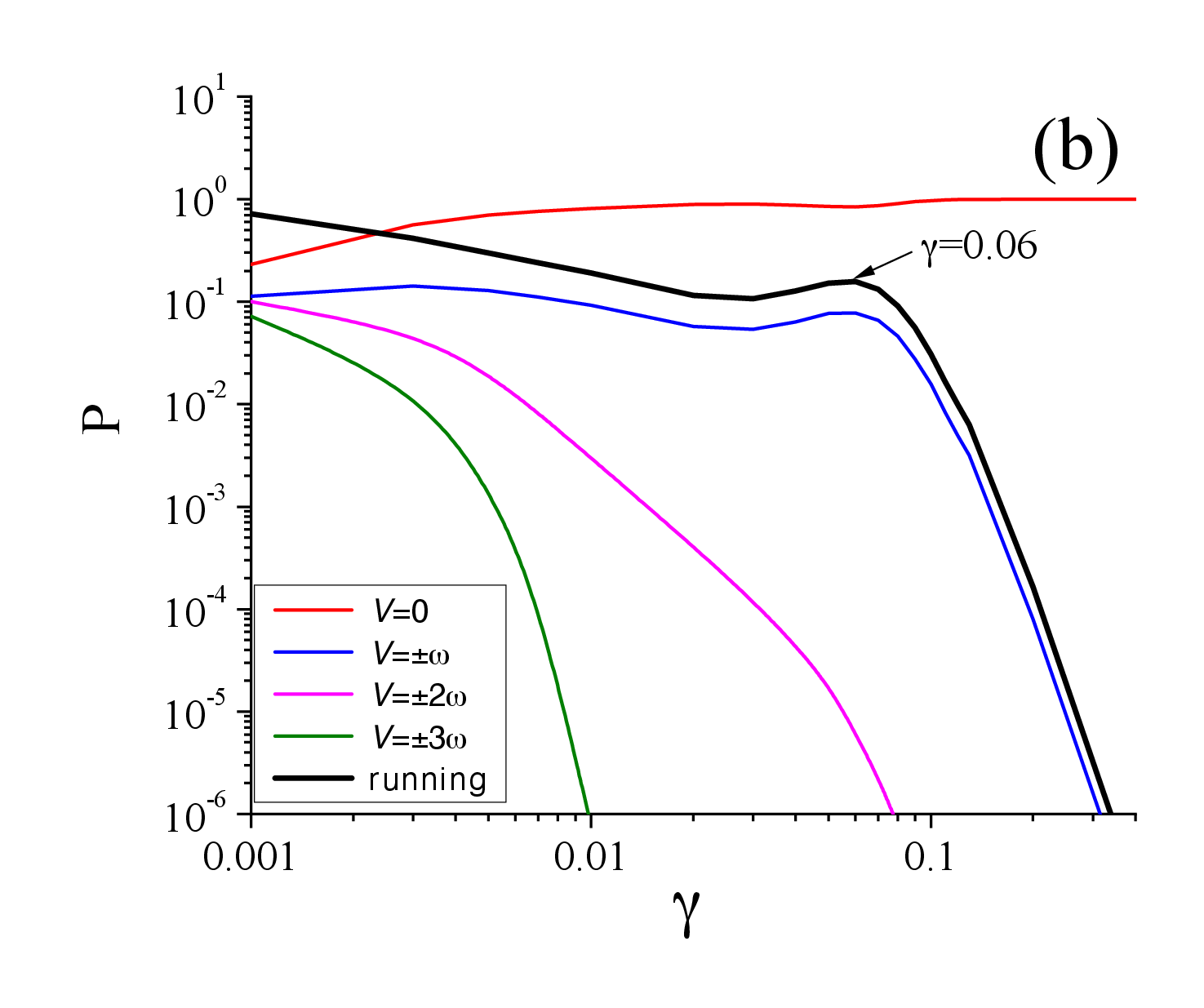}
\caption{The weak dissipation regime:  Panel (a) $D$ versus  $\gamma$.   Inset: The probability distribution $P(V)$ of  the period-averaged velocity $V$ of the Brownian particle. 
Panel (b) The stationary probability $\mbox{P}$ of the Brownian particle to reside in various  states $V = 0, V = \pm k\omega\}$ characterizing the deterministic counterpart of the studied setup. The black line corresponds to the sum of probabilities for all running states. Parameters are:  $a=14$, $\omega = 1.59$ 
and  $Q=0.05$. }
\label{fig4}
\end{figure} 

\ \ \ \\

\noindent {\bf B. Diffusion in weak dissipation regime} \\
 The non-monotonic dependence of diffusion on friction is also detected in the weak damping regime. In Fig. \ref{fig4} we depict this behavior. In the vicinity of $\gamma = 0.06$, the probability distribution $P(V)$ for the period-averaged velocity $V$ of the Brownian particle  has three remarkable maxima, one corresponding to the locked state $V=0$ and two symmetric running states of the averaged velocities $V=\pm\omega$. At non-zero temperature, there are random transitions induced by thermal fluctuations between these states. For one trajectory, the particle can follow the running solution, next for some time interval it resides in the localized state and later it again travels along the running state either in the same or the opposite direction. In this way, for a given trajectory various deterministic solutions are visited in a random sequence. In consequence the deterministic structure of locked and running states impacts the spread of trajectories. 

In panel (b) of Fig. \ref{fig4} we present the probability $\mbox{P}$ to reside in various states $\{V = 0, \pm k\omega\}$ characterizing the deterministic counterpart of the studied setup. It allows us to answer the question about the mechanism standing behind the observed pronounced maximum of the diffusion coefficient $D$ as a function of the friction $\gamma$ for weak damping regime $\gamma \approx 0.06$. The initial decay of this characteristics is clearly related to a decrease of the running states $V = \pm k\omega$ and simultaneous grow of the locked ones $V = 0$. Next, there is  one-to-one correspondence between the non-monotonic behavior of $D$ and the probability $\mbox{P}$ to reside in the running states. Therefore the local growth of the number of running trajectories is responsible for the increase of $D$ despite the fact that the damping $\gamma$ is getting bigger as well.  We note that this  mechanism of growing of $D$  is radically different than in the case of the strong dissipation regime.  

\begin{figure*}[t]
\centering
\includegraphics[width=0.4\linewidth]{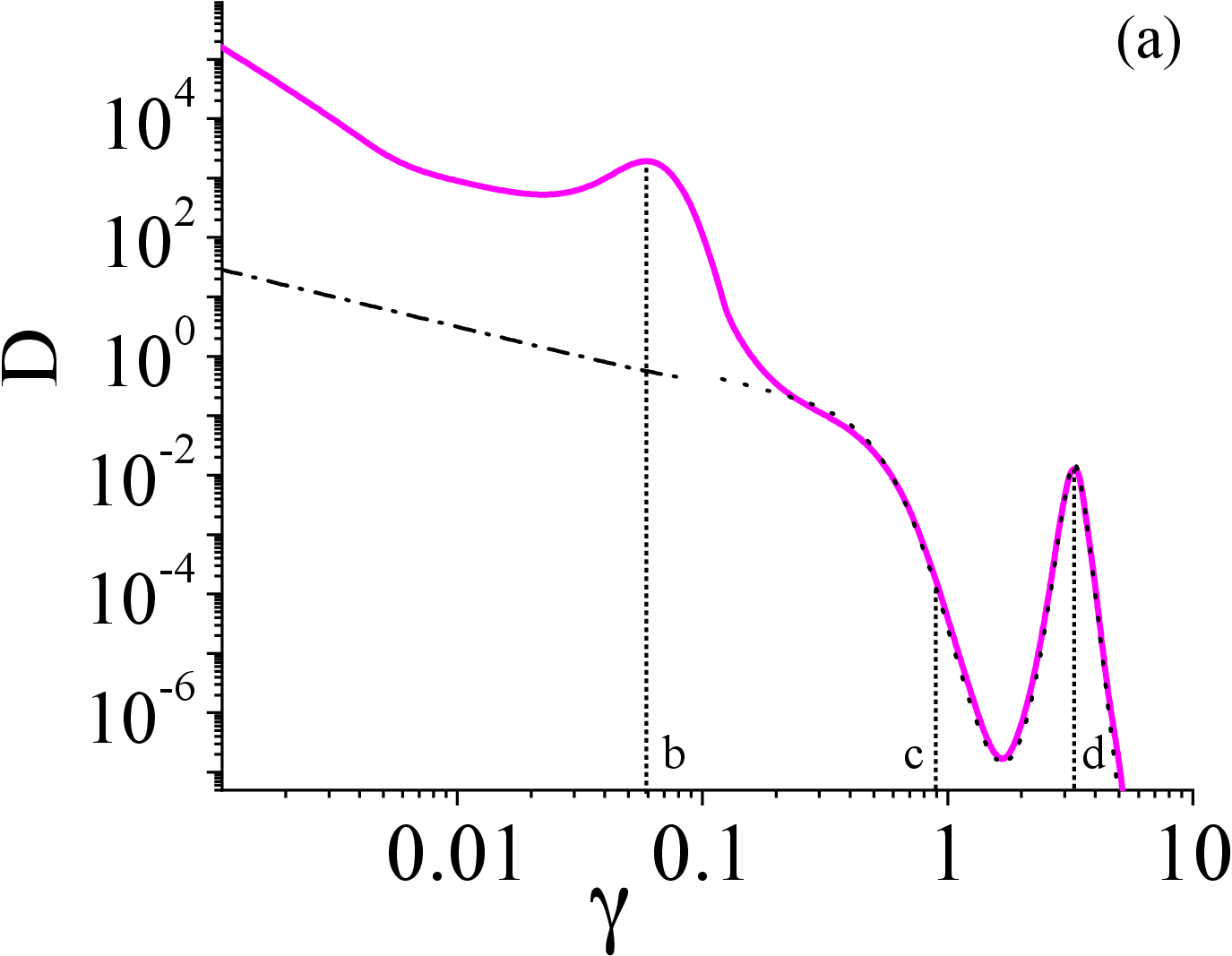} 
\includegraphics[width=0.4\linewidth]{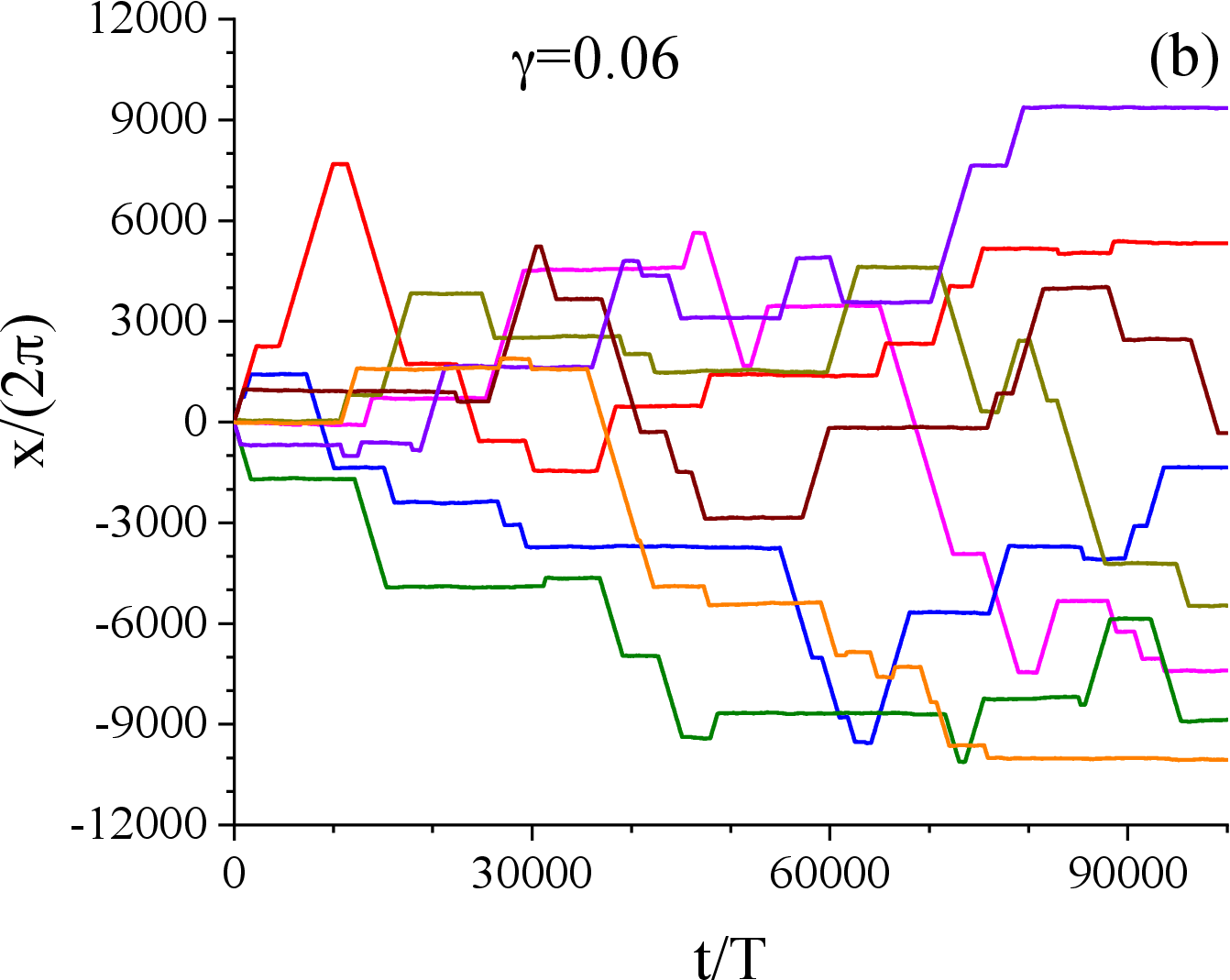}\\
\ \ \ \\

\includegraphics[width=0.4\linewidth]{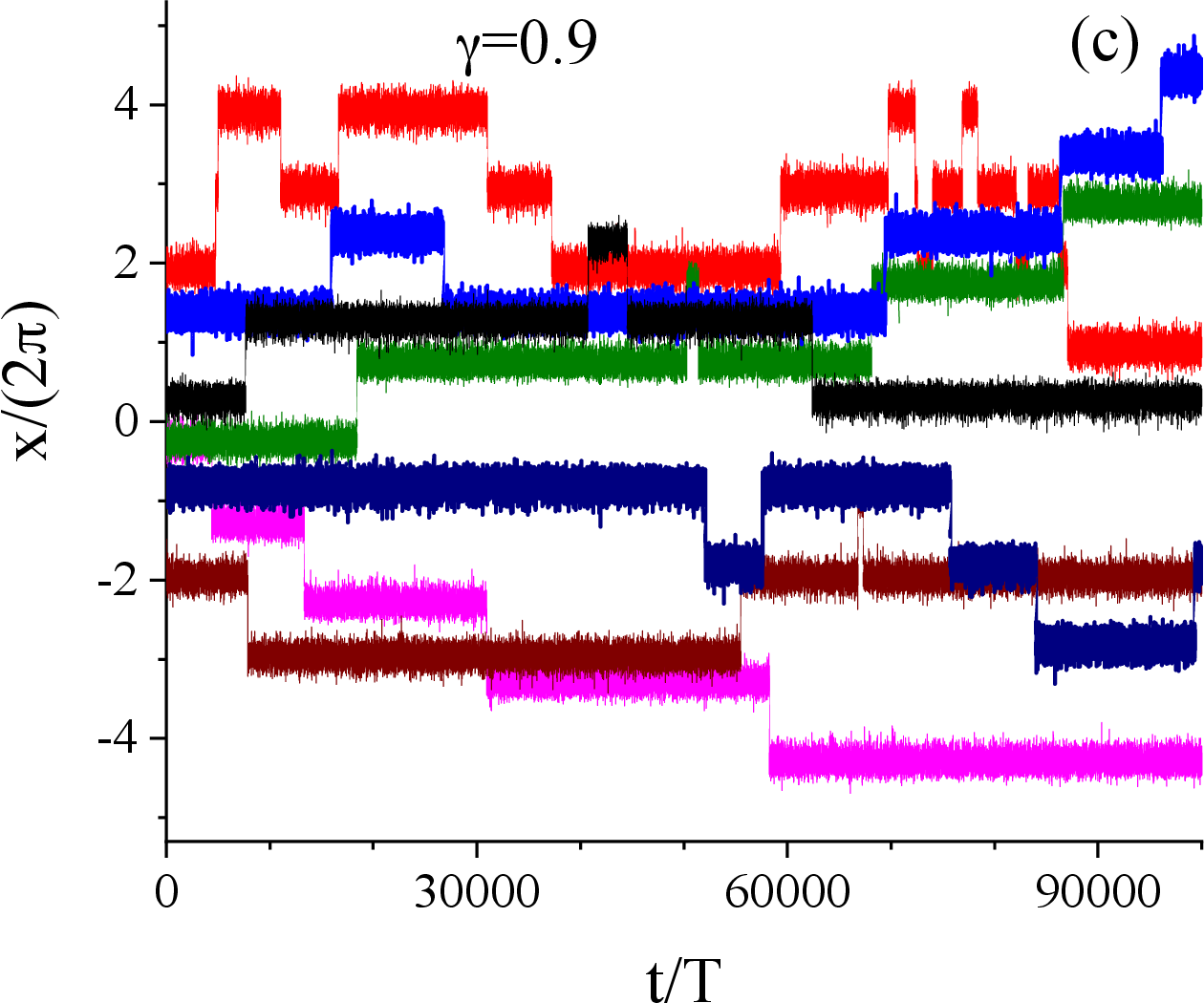}
	\includegraphics[width=0.4\linewidth]{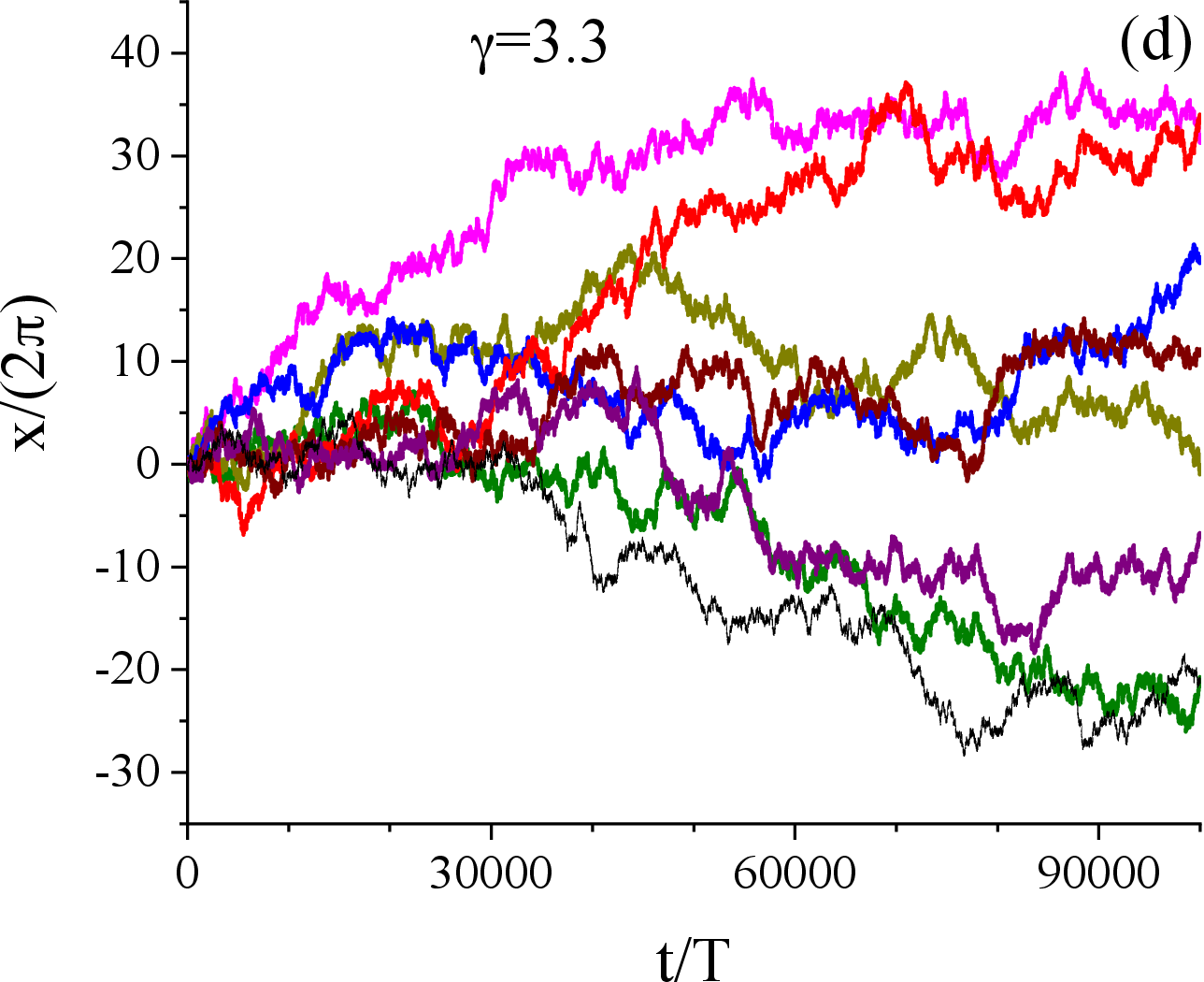}
	\caption{The diffusion coefficient $D$ versus  the friction coefficient $\gamma$.  Comparison of the  exact numerical  and   approximate  results given by Eqs. (\ref{strong}) (strong damping: dotted line) and (\ref{DMweak}) (weak damping: dotted-dashed line). The parameters are: $a=14$,  $\omega = 1.59$ and 
 $Q = 0.05$.    Exemplary set of the system trajectories vs. time $t/\mathsf{T}$ ($\mathsf{T} = 2\pi/\omega$) for $\gamma=0.06$, the first maximum (panel (b); $\gamma =1$, (panel c) and 
 $\gamma =3.3$, the second maximum (panel d).   Note that  the vertical axes have different scales  in the panels (b)-(d). }
\label{fig5}
\end{figure*}

In Fig. \ref{fig5}, we put together results of Fig. \ref{fig2} and Fig. \ref{fig4} and depict a spread of trajectories for three values of the friction $\gamma$. In panel (b), it is the weak dissipation regime with $\gamma=0.06$ (the first maximum of $D$) in which both running and locked trajectories emerge. In panel (c) $\gamma = 0.9$ and only locked trajectories exist. In panel (d), the strong dissipation regime with $\gamma = 3.3$ is presented for which there are only locked trajectories likewise. For this value of $\gamma$ the second maximum of $D$ is detected. We do not present a corresponding figure at the minimum $\gamma = 1.66$ of the diffusion coefficient $D$ due to the very small spread of trajectories.\\

\noindent  {\bf C. Vibrational mechanics approach}

It is argued that within vibrational mechanics methods \cite{sorokin} the high-frequency dynamics (\ref{La}) can be approximated by a much  simpler Langevin equation. Following the approach of  \cite{boroEPL,march-prl} we can separate
\begin{equation}
	\label{sep}
	x(t) = {\tilde x}(t) + \psi \sin(\omega t +\phi_1), 
\end{equation}
where  $\tilde x(t)$ represents a slowly time-modulated stochastic process. For $\omega \gg 1$, the fast oscillating terms can be averaged over the period of the ac driving and the Langevin equation (\ref{La})  for the slow reduced spatial
variable $\tilde x(t) \equiv \tilde x$ can be written as
\begin{equation}
	\label{VM}
	\ddot{\tilde x} + \gamma\dot{\tilde x} = -J_0(\psi) \sin{\tilde x} + \sqrt{2\gamma Q} \, \xi(t), 
\end{equation}
where  $J_0(\psi)$ is the zero order Bessel function \cite{specialF} of the argument  
\begin{equation}
	\label{psi}
	\psi = \psi(\gamma) = \frac{a}{\omega \sqrt{\omega^2 + \gamma^2}}. 
\end{equation}
 We now want to apply Eq. (\ref{VM}) to test its correctness in both strong and weak damping regimes for the case consider above, i.e. for the frequency $\omega =1.59$, which obviously does not fulfill the condition  $\omega \gg 1$.
We note that the starting total time-dependent potential $\mathbb{W}(x, t) = -\cos(x) -ax \sin(\omega t+\phi_0)$ is replaced by the time-independent renormalized potential $\mathbb{W}_{\mbox{eff}}(x) = -J_0(\psi) \cos(x)$. In this approach, the particle diffuses in a renormalized potential of the effective  
barrier $2J_0(\psi)$. In Fig. \ref{fig5} we compare results obtained from the exact dynamics (\ref{La}) and its approximate counterpart (\ref{VM}). 

For strong damping, the approximate expression for the diffusion coefficient can be obtained from Eqs.  (\ref{D1}) and (\ref{B1}),  which  takes now  the following form \cite{lifson,entropy23} 
\begin{equation}
	\label{strong}
	D = \frac{Q}{\gamma I_0^2(J_0(\psi)/Q)}, 
\end{equation}
where $I_0(x)$ is the modified Bessel function of the first kind \cite{specialF}.
As is seen in Fig. \ref{fig5},  the approximation (\ref{VM}) accurately reproduces the exact results for the friction coefficient $\gamma > 0.3$ which could be treated as the regime of strong damping. For the value $\gamma = 1.66$ it is a local minimum of $D$,  at this point  
$-J_0(\psi(1.66)) \approx 0.4$ and the effective potential barrier assumes the  maximal value. In turn, for $\gamma =3.3$ the diffusion coefficient is locally maximal and $J(\psi(3.3)) \approx 0$, i.e. the effective potential barrier dissapears. 
It turns out that for larger values of the amplitude $a$ new maxima emerge. For example,  if  $a=18$ there are two maxima at $\gamma_M = 1.05$ and $\gamma_M = 1.4$ but for $a=35$ four  maxima exist at $\gamma_M = 0.95, 2, 3.75, 9.5 $ (not shown). The envelope of these points behaves roughly like $D(\gamma_M) \sim 1/\gamma_M$.

For the case of weak dissipation, we can apply Eqs. (\ref{D1}) and (\ref{B2}) and then  \cite{pavli,entropy23}
\begin{equation}
	\label{DMweak}
	D =  \frac{Q}{\gamma}  \,  \frac{\sqrt{\pi d_0/ Q}}{ 2 I_0(d_0/Q)}  \int_{1}^{\infty}
	\frac{\mbox{e}^{-d_0 y/Q} \, dy}{\sqrt{y+1} \,  \mathbf{E}(\sqrt{2/(y+1)}) },
\end{equation}
where $d_0 = J_0(\psi)$ and  $\mathbf{E}(k)$ is the complete elliptic integral of the second kind \cite{specialF}.  In Fig. \ref{fig5} we compare the dependence $D(\gamma)$ from this relation with the exact results of simulations. In this regime the discrepancy is very large and (\ref{DMweak}) cannot reconstruct the local maximum for $a = 14$ which occurs at $\gamma = 0.06$. Therefore this approximation totally fails. It is due to the fact that the periodic part of the ansatz (\ref{sep}) takes into account only locked trajectories but not the running ones.
 In the regime of strong damping, solely locked solutions exist (see panel (d) in Fig. \ref{fig5}) and then the approximation  (\ref{sep}) is fine. On the other hand, in the regime of weak damping, there are also running trajectories  like in  Eq. (\ref{solut}), cf.  panel (b) in Fig. \ref{fig5}. Therefore the ansatz (\ref{sep}) fails and in consequence Eq. (\ref{VM}) does not reproduce properties of the starting system (\ref{La}). Its correct formulation is an interesting problem but lying beyond the scope of the current paper.
\\

\noindent {\Large {\bf Discussion} } \\
We showed that the dependence of the diffusion coefficient $D$ on the friction coefficient $\gamma$ characterizing the dissipation mechanism can render unexpected features. Typically,  $D$ is a decreasing function of $\gamma$. In contrast, we demonstrated that for an archetypal model of a nonequilibrium system consisting of a Brownian particle moving in a symmetric and spatially periodic potential subjected to an unbiased time-periodic driving the diffusion coefficient $D$ can be amplified up to several orders if the friction coefficient $\gamma$ is increased by a little bit. It can be realized for sufficiently large amplitude of the ac-driving and in the strong damping regime. 

We revealed the mechanisms standing behind this counterintuitive phenomenon which is related to   bifurcation of periodic orbits oscillating symmetrically around the potential maximum and their displacement towards the potential minimum when the friction coefficient increases. It induces the change of the effective barrier of the periodic potential and its vanishing at the maximum of $D$. Simultaneously, when $\gamma$ grows the intensity of thermal noise increases and, in turn, it intensifies  activation processes and transitions  between different states of the system. However, if 
damping is sufficiently strong, only decreasing of $D$ is observed as in the Einstein relation (1). 

Properties in the strong dissipation regime are very well reconstructed by the vibrational mechanics methods proposed in Refs. \cite{boroEPL,march-prl}.
On the other hand, in the weak dissipation regime, where the increase of diffusion vs friction  is also observed, the effect is generated by a non-monotonic change of population of the running orbits. However, in this regime the enhancement of diffusion is small. Moreover, the vibrational mechanics approximation in the form (\ref{sep})-(\ref{VM})  ceases to function in this regime.

We discussed only a single set of parameters  $\{a, \omega, Q\}$ for which the giant acceleration of the diffusion process is detected. Are these three values exceptional? To answer this question we have searched the part of the parameter space in order to locate regions of giant enhancement of $D$. The inspection of the results revealed that there are many sets of parameter values for which this behavior occurs.

The phenomenon of giant enhancement of diffusion has been studied both theoretically and experimentally in a  number of systems. In Ref. \cite{peter-epl}  the overdamped Brownian particle moving in a piecewise linear spatially periodic potential and driven by a piecewise constant time periodic driving was considered. It was shown that the diffusion coefficient exhibits quasi-periodic behavior as a function of the "tilting time" of the piecewise constant force and at the maxima $D/D_E$ can be much larger than 1. In Refs. \cite{reimann2001a, reimann2001b} the overdamped dynamics of the Brownian particle in the washboard potential was studied and it was shown analytically that in a critically biased periodic potential  the diffusion coefficient as a function of a constant force  may be, in principle, arbitrarily enhanced. For a realistic experimental setup, an enhancement by 14 orders of magnitude was  predicted so that thermal diffusion could  be observable on a macroscopic scale at room temperature. The extension to the  underdamped regime has been presented in Refs. \cite{marchenko2012,lindner16} and the regions of parameters for which the diffusion coefficient exponentially grows with inverse temperature have been identified. In Ref. \cite{urbakh}  it was  demonstrated that lateral vibrations of the substrate can dramatically increase surface
diffusivity  and reduce friction at the nanoscale. It was analyzed in the context of  atomic force microscopy  by using the model of a tip interacting with a substrate, which oscillates in the lateral direction. The motion of the tip in the   lateral, and normal, directions is governed by the coupled Langevin equations and it was  found a sharp transition from the state with  a small tip-surface separation to the state with a large separation as the vibration frequency increases. 

The giant diffusion has been confirmed by experiments. In Ref. \cite{grier}, a single colloidal sphere circulating around a periodically modulated optical vortex trap exhibits normal diffusion of the Eisnstein-like law with  an effective diffusion coefficient that is enhanced by more than 2 orders of magnitude. In the next experiment \cite{nano}, the diffusion of a gold nanoparticle in the nonconservative force field of an optical vortex lattice was observed and radiation pressure in the vortex array is shown to induce an  enhancement by 2 orders of magnitude over the free thermal diffusion. In Ref. \cite{hayashi},  diffusion to a single-molecule experiment on a rotary motor protein, $F_1$-ATPase , which is a component of $F_0F_1$ adenosine triphosphate synthase, was investigating. The rotation of $F_1$ was induced by   applying an external torque using a single-molecule technique. It was found that the diffusion coefficient of a probe attached to $F_1$ (as a function of the external torque) shows a resonance peak, which was predicted by the theory of the giant acceleration  of diffusion in Refs. \cite{reimann2001a, reimann2001b}. In the recent experiment \cite{rotor},  the setup consists of a vacuum optical tweezer trapping a dielectric silica nanodumbbell. Its motion is driven by an elliptically polarized light beam tilting the angular potential. By varying the gas pressure (i.e. the friction) around the point of maximum intermittency, the  effective diffusion coefficient increases by more than 3 orders of magnitude over free-space diffusion.

Properties of the studied system could be corroborated experimentally in several setups. We would like to mention two of them. The first is a resistively and capacitively shunted Josephson junction device. Dynamics of the phase difference $\Psi=\Psi(t)$ between the macroscopic wave functions across the Josephson junction in the framework of the Stewart-McCumber model \cite{stewart,mccumber,kautz1996} is described by the following equation
\begin{eqnarray} \label{JJ1}
\Big( \frac{\hbar}{2e} \Big)^2 C\:\ddot{\Psi} + \Big( \frac{\hbar}{2e} \Big)^2 \frac{1}{R} \dot{\Psi} + \frac{\hbar}{2e} I_0 \sin \Psi =  \frac{\hbar}{2e} I(t) + \frac{\hbar}{2e}
\sqrt{\frac{2 k_B T}{R}} \:\xi (t) .
\end{eqnarray}
The left hand side contains three additive contributions: a displacement current due to
the capacitance $C$ of the junction, a normal (Ohmic) current characterized
by the normal state resistance $R$ and a Cooper pair tunnel current characterized by
the critical value $I_0$. In the right hand side, $I(t)$ is an external current applied to the device. 
There is an evident correspondence between the dimensional Langevin equation  (\ref{model1}) (see below) and Eq. (\ref{JJ1}): the coordinate $x=\Psi$,    
the mass $M=(\hbar /2e)^2 C$,  the friction coefficient $\Gamma = (\hbar/2 e)^2(1/R)$, the barrier height $\Delta U = (\hbar/2 e) I_0$ and the period $L=2\pi$.  Note that the increase of the friction $\Gamma$ now means the decrease of the resistance $R$. 
The particle velocity $v=\dot x$ corresponds to the voltage drop $V$ across the junction and  the generalized Green-Kubo relation for the periodically driven processes allows to investigate the diffusion coefficient, see Appendix in Ref. \cite{jpc}. 
Moreover, in this context it is worth mentioning the recent experiments concerning measurements of phase dynamics in Josephson Junctions and SQUIDs \cite{phase} which open another possibility to test our findings. 

The second example is related to an experimental setup based on the optical trapping of a nanodumbbell in a moderate vacuum. The system is trapped in an elliptically polarized laser beam \cite{rotor}. For linearly polarized optical tweezers the dumbbell is subjected to a periodic force $\sin(x)$, cf. Eq. (1) in Ref. \cite{rotor}.  A special feature of this experiment is that the friction coefficient being linearly proportional to the gas pressure can be tuned over several orders of magnitude. It makes this optical trapping setup a suitable platform to study stochastic dynamics in the low-friction and strong-friction regimes. Recently, an effective technique called the optical feedback trap has been proposed \cite{feedback}. It allows to create, with a high precision position detection and ultrafast feedback control, a mathematically driven spatio-temporal effective potential of any desired shape using the optical feedback force. 

Our results could be applied, for instance, in microfluidic devices for controlled separation of particles through diffusion \cite{separator}. The main idea is the following. There exist two species of particles with the linear sizes $R_1$ and $R_2=2 R_1$. 
From the Stokes relation, $\gamma \propto R$, the corresponding friction coefficients for them are $\gamma_1$ and $\gamma_2=2\gamma_1$, and the diffusion coefficients are $D_1$ and $D_2$, respectively. If $\gamma_1 \approx 1.65$ (see Fig. 2) then $\gamma_2 \approx 3.3$ and $D_2 \approx 5.6 \times 10^4 \, D_1$. From the relation (\ref{D}) it follows that, statistically, if the smaller particles overcome the distance $X_1$ then the larger ones overcome the distance $X_2 \approx \sqrt{5.6 \times 10^4} \, X_1 = 2.4 \times 10^2 \, X_1$, i.e. two orders of magnitude greater. If the separator is properly designed, the larger particles can be effectively separated from the smaller ones. Recent developments of size-selective separation methods for nanoparticles using the external fields, filtration, and the stability of the colloid system are reviewed in Ref. \cite{mori}.\\

\noindent {\Large {\bf Methods} }\\
\noindent {\bf  Dimensional Langevin equation and scaling } \\
The dimensionless Langevin equation (\ref{La}) is obtained from the dimensional Langevin equation in the  form \cite{entropy23,marchPRE23,spiechowicz2016njp}
\begin{equation}
	\label{model1}
	M\ddot{x} + \Gamma\dot{x} = -U'(x) + F(t) + \sqrt{2\Gamma k_B T}\,\xi(t),
\end{equation} 
where  $M$ is the mass of the Brownian particle, $U(x)$ is the  spatially periodic potential $U(x)=U(x+L)$ of period $L$  and $F(t)$ is  an unbiased and symmetric time-periodic force. Moreover,  the dot and the prime denotes  differentiation with respect to time $t$ and the particle coordinate $x$, respectively. The parameter $\Gamma$ is the friction (damping) coefficient. The potential $U(x)$ has the simplest form 
\begin{equation}
	\label{potential}
	U(x) =  -\Delta U\cos{\left( \frac{2\pi}{L}x \right)}, 
	\end{equation}
	where $2 \Delta U$  is the  barrier height. 
The external ac-driving force of amplitude $A$ and angular frequency $\Omega$ is taken in the  form  
\begin{equation}
	F(t) = A \sin{(\Omega t + \phi_0)}, 	
\end{equation}
where $\phi_0$ is the initial phase. 
Thermal equilibrium fluctuations describing the influence of thermostat of temperature $T$ are normally assumed as  $\delta$-correlated Gaussian white noise of zero-mean value, 
\begin{equation}
	\langle \xi(t) \rangle = 0, \quad \langle \xi(t)\xi(s) \rangle = \delta(t - s), 
\end{equation}
where the bracket $\langle \cdot \rangle$ denotes an average over noise realizations (ensemble average). 
The noise intensity $2\Gamma k_B T$ in Eq. (\ref{model1}) follows from the fluctuation-dissipation theorem \cite{kubo1966}, where $k_B$ is the Boltzmann constant. 

We define the following dimensionless coordinate and time 
\begin{equation}
	\label{scaling}
	\hat{x} = 2\pi \frac{x}{L}, \quad \hat{t} = \frac{t}{\tau_0}, \quad \tau_0 = \frac{L}{2\pi}\sqrt{\frac{M}{\Delta U}}.
\end{equation}
The characteristic length $L$ is the period of the potential $U(x)$  and the characteristic time $\tau_0$ is related to the period of small oscillations inside the  well  of $U(x)$. 

Under the above scaling, Eq. (\ref{model1}) is converted to the form
\begin{equation}
	\label{dimless}
	\ddot{\hat{x}} + \gamma\dot{\hat{x}} = -\sin{\hat{x}} + a \sin (\omega \hat{t} + \phi_0) +  \sqrt{2\gamma Q} \, \hat{\xi}(\hat{t}), 
\end{equation}
where the  dimensionless parameters read  
\begin{eqnarray}
\label{parameters}
	\gamma =  \frac{\Gamma L}{2\pi \sqrt{M\Delta U}}, \quad a = \frac{1}{2\pi}\frac{L}{\Delta U} A, 
\quad \omega =   \Omega \frac{L}{2\pi}\sqrt{\frac{M}{\Delta U}}, \quad Q = \frac{k_B T}{\Delta U}.  
\end{eqnarray}
The dimensionless  potential $W(\hat x)$ of the period $L=2\pi$ is  $W(\hat{x}) = U((L/2\pi)\hat{x})/\Delta U = -\cos{\hat x}$ and the corresponding potential force reads  $f(\hat x) = -W'(\hat{x})=-\sin \hat{x}$. The rescaled thermal noise is $\hat{\xi}(\hat{t}) = (L/2\pi \Delta U)\xi(t) = (L/2\pi \Delta U)\xi(\tau_0\hat{t})$ and has the same statistical properties as $\xi(t)$; i.e., $\langle \hat{\xi}(\hat{t}) \rangle = 0$ and $\langle \hat{\xi}(\hat{t})\hat{\xi}(\hat{s}) \rangle = \delta(\hat{t} - \hat{s})$. The dimensionless noise intensity  $Q$ is the ratio of thermal energy and half of the activation energy the particle needs to overcome the nonrescaled potential barrier.  In order to simplify the notation  we omit the hat-notation in Eq. (\ref{La}). \\

The main problem studied in the paper is the dependence of the dimensionless diffusion coefficient $D$ on the dimensionless friction coefficient $\gamma$, see Eq. (\ref{parameters}). 
If we want to consider a regime of the strong damping $\gamma \gg 1$ then it can be realized in two ways: (i) for fixed $M$ the friction coeffcient $\Gamma$ is large or (ii) for fixed $\Gamma$ the mass $M$ is small. If the control parameter is the dimensional friction coefficient $\Gamma$ then from Eq. (\ref{parameters}) it follows that it is equivalent to the change of the dimensionless $\gamma$ while other parameters are fixed. However, if the control parameter is the dimensional mass $M$ then from Eq. (\ref{parameters}) it follows that both $\gamma$ and $\omega$ are changed simultaneously. Therefore these two scenarios leading to the regime of strong damping are not equivalent if the remaining parameters are supposed to be fixed.  
\\

\noindent {\bf Diffusion coefficient: Selected analytical results}   \\
An analytical expression for the diffusion coefficient $D$ is here  presented for some selected models.  For the force-free Brownian particle or under the influence of a constant force $F$, i.e. when Eq.(\ref{La}) assumes the form 
\begin{equation}
	\label{L2}
	\ddot{x} + \gamma\dot{x} =  F +  \sqrt{2\gamma Q} \, \xi(t), 
\end{equation}
the diffusion coefficient it given by the Einstein relation \cite{entropy23}
\begin{equation}
	\label{D0}
	D_E =  \frac{Q}{\gamma}.
\end{equation}
If the Brownian particle moves in random medium and this problem is modeled in terms of the fluctuating  viscous friction 
\begin{equation}
	\label{gama}
	\gamma \to  \gamma(t) =\gamma_0 + z(t) > 0, 
\end{equation}
where  fluctuations $z(t)$ are assumed to be the zero-mean stationary random process  $\langle z(t)\rangle =0$, 
then the diffusion coefficient is given by the expression \cite{Talkner1,Talkner2} 
\begin{equation}
	\label{DG}
	\tilde D =  Q \int_0^{\infty} ds \, \mbox{e}^{-\gamma_0 s} \, 
	\left\langle \exp\left(-\int_0^s du \, z(u) \right) \right\rangle 
	\ge \, D_E, 
\end{equation}
where averaging on the right-hand side has to be performed over all realizations of fluctuations $z(t)$. Applying  the Jensen inequality  $\langle \exp(\xi)\rangle \ge \exp(\langle \xi \rangle )$, we note that $\tilde D$  is always greater than the Einstein diffusion coefficient. 

For the Brownian particle moving in the periodic structure, i.e. when the Langevin equation reads 
\begin{equation}
	\label{L3}
	\ddot{x} + \gamma\dot{x} = -d_0 \sin{x} +  \sqrt{2\gamma Q} \, \xi(t), 
\end{equation}
the corresponding diffusion coefficient is derived in two limiting damping regimes in which it can be expressed as  
\begin{equation}
	\label{D1}
	D =   \frac{Q}{\gamma} \, B(Q). 
\end{equation}
 In the case of overdamped dynamics ($\gamma \gg 1$), i.e. when the inertial term in Eq. (\ref{L3}) is formally neglected, it assumes the following form \cite{lifson,entropy23}
\begin{equation}
	\label{B1}
	B(Q) = \frac{1}{I_0^2(d_0/Q)}. 
\end{equation}
In the underdamped regime ($\gamma \ll 1$), the factor $B(Q)$ reads \cite{pavli,entropy23}
\begin{equation}
	\label{B2}
	B(Q)=  \frac{\sqrt{\pi d_0/ Q}}{ 2 I_0(d_0/Q)} \int_{1}^{\infty} \frac{\mbox{e}^{-d_0 y/Q} \, dy}{\sqrt{y+1} \,  \mathbf{E}(\sqrt{2/(y+1)}) },
\end{equation}
where $I_0(x)$ is the modified Bessel function of the first kind and $\mathbf{E}(k)$ is the complete elliptic integral of the second kind \cite{specialF}. 
In both cases $B(Q)$ depends only on $d_0$ and the dimensionless temperature $Q$.  If $d_0$ does not depend on other parameters of the system then 
the diffusion coefficient (\ref{D1}) is smaller than the Einstein diffusion coefficient (\ref{D0}), i.e. $D \le D_E$.  
  Eq. (\ref{L3}) has the same form as Eq. (\ref{VM}) in which $d_0 = J_0(\psi)$ and $\psi = \psi(a, \omega, \gamma)$. Therefore dependence of $D$ 
  on $\gamma$ is more complicated than in Eq. (\ref{D0}). We stress that Eq. (\ref{L3})  is  applied for  verification of approximations in the strong and weak dissipation regimes, see subsection:  Vibrational mechanics approach. \\
 
\noindent {\bf Data availability} \\
The datasets generated during the current study are available from the corresponding author on reasonable request.


\vspace{10pt}

\noindent {\bf Acknowledgments} \\
This work was supported by the Grants NCN 2022/45/B/ST3/02619 (J.S.)  and NASU No.  0122U002145/2022-2023 (V.A). I. G. M. acknowledges University of Silesia for its hospitality since the beginning of the war, 24 February 2022. \\
{\bf Author contributions} \\ 
Conceptualization, discussion, writing and editing of the manuscript: I.G.M., J.{\L.} and J.S.; simulations and calculations:  I.G.M, V.A., I.I.M. and J.S. All authors have read and agreed to the published version of the manuscript. \\
{\bf Competing interests } \\
The authors declare no competing interest  \\
{\bf Additional information } \\
Correspondence and requests for materials should be addressed to J.Ł. (e-mail: jerzy.luczka@us.edu.pl) 

\end{document}